# Enhanced UV light detection using wavelength-shifting properties of Silicon nanoparticles

S. Magill,[a,*] M. Nayfeh,[b] M. Fizari,[b] J. Malloy,[b] Y. Maximenko,[b] J. Xie[a] and H. Yu[b]

[a] *Argonne National Laboratory,*
  *Argonne, Illinois, USA*

[b] *University of Illinois at Urbana-Champaign,*
  *Urbana, Illinois, USA*
  *E-mail*: srm@anl.gov

*ABSTRACT:* Detection of UV photons is becoming increasingly necessary with the use of noble gases and liquids in elementary particle experiments. Cerenkov light in crystals and glasses, scintillation light in neutrino, dark matter, and rare decay experiments all require sensitivity to UV photons. New sensor materials are needed that can directly detect UV photons and/or absorb UV photons and re-emit light in the visible range measurable by existing photosensors. It has been shown that silicon nanoparticles are sensitive to UV light in a wavelength range around ~200 nm. UV light is absorbed and re-emitted at wavelengths in the visible range depending on the size of the nanoparticles. Initial tests of the wavelength-shifting properties of silicon nanoparticles are presented here that indicate by placing a film of nanoparticles in front of a standard visible-wavelength detecting photosensor, the response of the sensor is significantly enhanced at wavelengths < 320 nm.

---

[*] Corresponding author.

# Contents

# 1. Introduction

With the development of clear, dense crystals for HEP applications [1], a homogeneous calorimeter made entirely of crystal could contain high-energy jets in a detector volume similar to those in existing experiments. Using Dual Readout capability (simultaneous detection of both Cerenkov and scintillator light), unprecedented energy resolution for hadrons produced in $e^+e^-$ interactions has been demonstrated in simulations [2]. It is anticipated that the amount of (visible wavelength range) scintillation light produced by charged particles in the crystal will be more than is needed, so a few small photosensors (e.g., SiPMs) located on a crystal face or at corners could be used to collect that component while maintaining a high degree of homogeneity. For the Cerenkov light, however, it is foreseen that light collection will be required on several sides of the crystals with a collection area comparable to that of the crystal face. A photosensor is needed which 1) must be able to detect photons in the UV wavelength range (< 350 nm), 2) must be thin so as to preserve the homogeneity of the calorimeter, 3) must be of large area (~few $cm^2$), and 4) should be blind to visible light so as not to detect light from scintillation.

It has been shown that Si nanoparticles change the photon detection properties of the element, shifting its response from the visible wavelength range into the UV [3]. Apparently, the band gap of Si in nanoparticle form is increased to over 3 eV, which can be compared to its normal value in elemental form of 1.1 eV. In addition, by choosing the size of the nanoparticles, re-emission of wavelength-shifted light can be tuned - 1 nm nanoparticles produce blue light while 3 nm nanoparticles produce red light. Furthermore, when manufactured into a photodiode by depositing a thin layer of Si nanoparticles on a p-type silicon substrate, the response shifted into the UV range at ~200 nm and was essentially zero for wavelengths > 400 nm. This approach would seem to satisfy several of the requirements for detection of Cerenkov light in a crystal calorimeter, so tests were initiated to confirm the results from [3].

## 2. Preparation of Si nanoparticle samples

The nanoparticles are prepared from Si wafers by chemical etching in HF and $H_2O_2$ using electrical or hexachloroplatinic acid catalyst. The nanoparticles are recovered from the treated wafer into a solvent of choice, such as isopropyl alcohol using ultrasound. We produce discrete size $Si_nH_x$ particles that are 1.0 ($Si_{29}H_{24}$), 1.67 ($Si_{123}$), 2.15, and 2.95 nm in diameter [4-5]. The particles of all sizes absorb efficiently in the UV range 3-10 eV, with strong novel luminescence in the visible with an optical band gap hence spectral distribution that depends on the size of the particle. The mono dispersed nanoparticles may subsequently be delivered from the liquid on a target using several delivery procedures, including electrospray, atomization, spin coating, or drop-drying. Quantum Monte Carlo (QMC) theory and simulation of the 1nm particle, which employ Hartree-Fock pseudopotentials, yields a simulated prototype of $Si_{29}H_{24}$ which is depicted in Fig. 1 (Left) [6]. Fig. 1 (Right) is a transmission electron microscope (TEM) image of 3 nm particles.

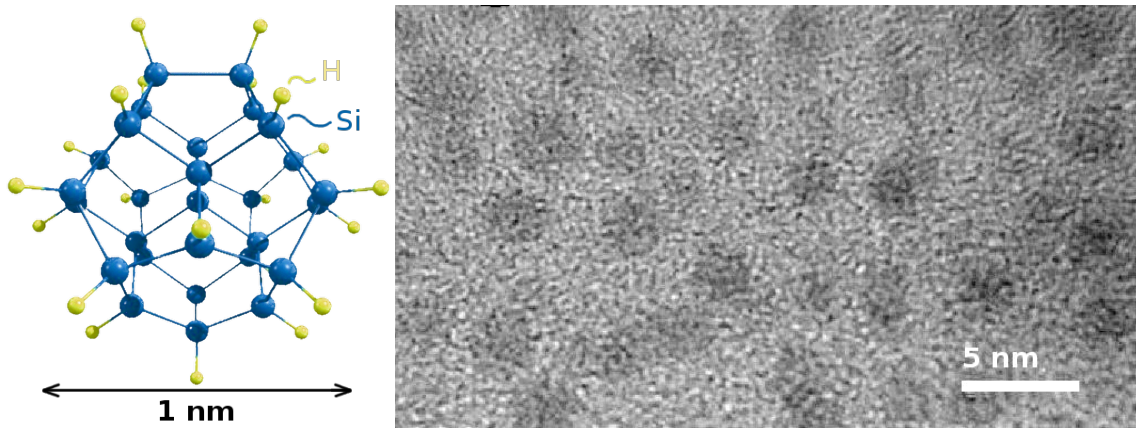

Figure 1. (Left) Simulated prototype of 1 nm Si nanoparticles, and (Right) TEM figure of 3 nm Si nanoparticles.

## 3. Establishment of baseline response

As a first test to verify that Si nanoparticles are sensitive to UV light, a thin layer of nanoparticles was deposited on a clear plastic film (polypropylene), which was placed on top/in front of the active surface of a photosensor. The deposit was done on the film instead of directly onto the photosensor to avoid damage to the resin coating during the nanoparticle deposition process and to allow flexibility in the testing process. In this test we used 3 nm diameter Si nanoparticles, which were suspended in isopropyl alcohol - several drops of the nanoparticle liquid solution were smeared on the plastic film and the alcohol was allowed to evaporate in room temperature air. No attempt was made to ensure that the nanoparticle layer was uniform or of any determined thickness. After drying, the presence of nanoparticles on the plastic film was confirmed by shining UV light on the coated film and observing a faint reddish glow with the naked eye. Under UV irradiation, Si nanoparticles of ~3 nm in diameter re-emit light in the range 550-750 nm with a peak intensity at ~650 nm as shown in Figure 2.

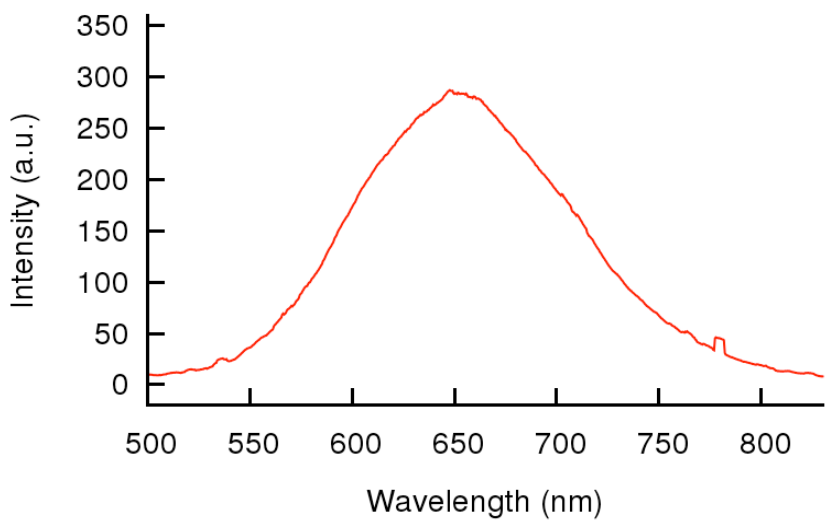

Figure 2. Luminescence of 3 nm nanoparticles under irradiation with 365 nm light.

### 3.1 Photosensor and monochromator details

The photosensor used was a Hamamatsu MPPC, which is a Silicon PhotoMultiplier (SiPM) of 3 mm x 3 mm square area with 50 micron pixel pitch mounted in a ceramic enclosure [7]. The photon detection efficiency is ~10% for 320 nm light, rising to a maximum of ~37% at 450 nm, then falling off to ~5% near 900 nm in wavelength. The same SiPM was used in all of the tests reported here. The match of 3 nm diameter Si nanoparticles with this particular SiPM is not ideal, since the peak emission of the nanoparticles is at a wavelength (~650 nm) not optimal for detection by the SiPM. Future tests will be done with 1 nm size Si nanoparticles, which emit light at a peak wavelength that closely corresponds to the maximum sensitivity peak of the SiPM.

The tests were performed on a fully automated optical-electrical measurement system consisting of a Newport 30 W continuum deuterium ($D_2$) lamp, a Newport Apex monochromator, a low noise Femto DLPCA-200 current amplifier and a Keithley 2701 multimeter. A detection range of ~160 nm to 900 nm can be achieved with the above system. The $D_2$ lamp intensity is a smoothly-falling function of increasing wavelength with a few narrow spikes in intensity at wavelengths of 490 nm, 580 nm, and 660 nm. More details about this system can be found in reference [8].

### 3.2 Baseline response

To establish a baseline response, the SiPM was illuminated with the $D_2$ lamp. The SiPM was connected to the current amplifier with its adjustable gain set to the maximum value of $10^9$. No bias voltage was applied to the SiPM - these tests were run with the sensor operating in photovoltaic mode. Several scans were made in wavelength steps of 5 or 10 nm depending on the range of the scan. The amplified current signal was displayed on the multimeter and recorded on a PC. The current signal obtained is the product of the wavelength-dependent response of the SiPM and the power spectrum of the $D_2$ lamp. This response is shown in Figure 3.

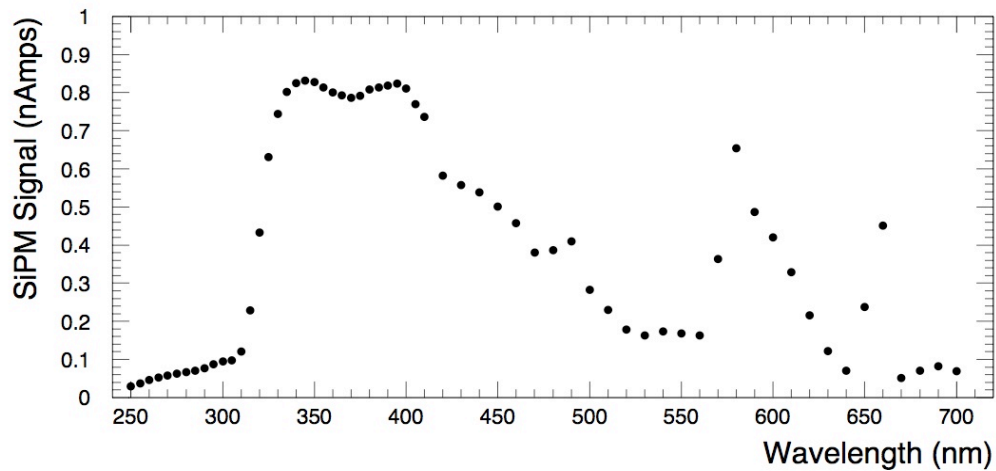

Figure 3. Current response of the SiPM in the wavelength range 250 - 700 nm.

The response below ~320 nm increases slightly with wavelength, then there is a sharp turn-on at ~320 nm, then a broad peak from 340 - 450 nm dominated by the SiPM photon detection efficiency in the region of relatively smoothly-falling $D_2$ lamp intensity, then the presence of several strong peaks from the $D_2$ lamp (490 nm, 580 nm, and 660 nm).

## 4. Experimental procedure

Several runs were taken of the light through the plastic film with no nanoparticle coating (Uncoated) and through the plastic film with the nanoparticle coating in 2 configurations - 1) with the coating on the $D_2$ lamp side of the plastic film (Coated), and 2) with the coating on the SiPM side of the plastic (Coated-Flipped). Figure 4 is a (not to scale) schematic diagram of these 3 configurations.

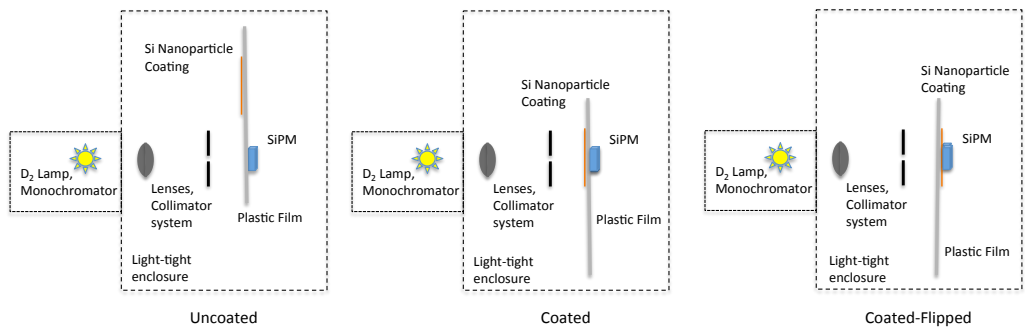

Figure 4. (Left) Uncoated configuration – light passes through plastic film only to SiPM, (Center) Coated configuration – light passes through Si nanoparticle coating then plastic film to SiPM, and (Right) Coated-Flipped configuration – light passes through plastic film then Si nanoparticle coating to SiPM.

The lenses and collimator system focused and formed the light into a 1 mm diameter spot at the surface of the SiPM, which has an active area of ~9 mm$^2$. The Si nanoparticle coating is ~1 cm in diameter and of undetermined and non-uniform thickness. Since the plastic is ~opaque to UV light, no low-wavelength signal was expected from the plastic film only and from the Coated-Flipped configuration. It was expected that the Coated configuration should show enhancement since the incident UV light (which can not penetrate the plastic) should be absorbed by the nanoparticle coating and re-emitted at ~650 nm which then passes through the plastic to be detected by the SiPM.

**5. Systematic Effects**

Light beam alignment from the $D_2$ lamp to the SiPM was eliminated as a systematic effect by fixing the SiPM with respect to the light beam spot. The beam spot size was forced by focusing and collimation to be ~1 mm in diameter and the SiPM was aligned so that the beam spot was centered on the SiPM active area. No stray light could enter the SiPM from either outside the light-tight enclosure or scattered light from the $D_2$ lamp. The size of the Si nanoparticle coating was ~1 cm in diameter, however, the thickness was not controlled over the coating area and was not determined. In the Coated and Coated-Flipped case, the position of the nanoparticle coating was arranged so that the same approximate area was seen by the SiPM in each case. In future tests, both the nanoparticle coating thickness and its uniformity will be determined.

Several runs were taken for the SiPM response with no plastic film and for the response through the plastic (Uncoated) in order to determine the variation in response over the time of the tests. In both cases, the variation seen was ~2% (1.9% for no plastic film and 2.1% for the Uncoated case).

**6. Results**

The first run was taken in 5 nm wavelength steps from 250 nm - 400 nm. In this test, response from plastic-only propagation (Uncoated) was compared to the response with the coating preceding the plastic film (Coated) and is shown in Figure 5.

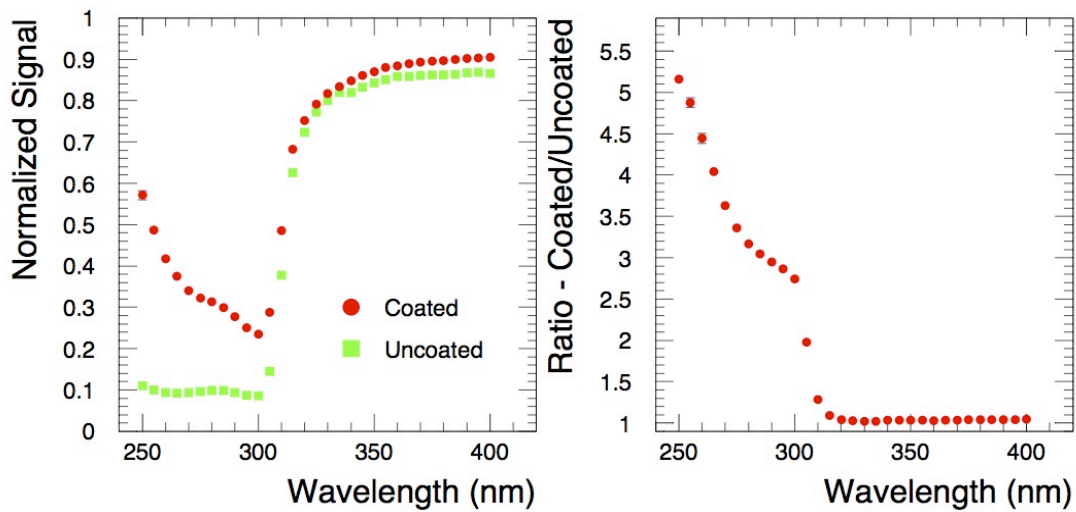

Figure 5. (Left) Response of SiPM through uncoated plastic film (squares), and through plastic film with Si nanoparticle coating (dots), and (Right) Ratio of coated to uncoated signals.

In both cases, the response as shown in Figure 5 was normalized by the SiPM response to the $D_2$ lamp as shown in Figure 3. In this case, the relative response between the Coated and Uncoated configurations is seen as evidence for the wavelength-shifting properties of the Si nanoparticles.

– 5 –

An increase in the response of the SiPM for decreasing wavelengths less than 320 nm is seen in the Coated response in contrast to the decreasing response of the SiPM only (Fig. 3) and the flat response of the Uncoated signal. At the lowest wavelength tested here (250 nm), the Coated response is a factor of ~5 greater than the Uncoated response. Recall that the wavelength has been shifted to ~650 nm - not optimal for this SiPM. Note that a significant rise in response is seen at the lower limit of the test (250 nm) - this is where the maximum in response of the first Si nanoparticle photodiode was seen [3].

A second test was done to measure the SiPM response down to the limit of the monochromator-$D_2$ lamp setup. With the fused silica window on the $D_2$ lamp, the lowest obtainable wavelength is ~150 nm. For lower wavelengths, a MgF window on the lamp and propagation through vacuum is required. Figure 6 shows the response of the SiPM in 5 nm wavelength steps from 150 nm to 400 nm, comparing the response of the Coated configuration to that of the Coated-Flipped.

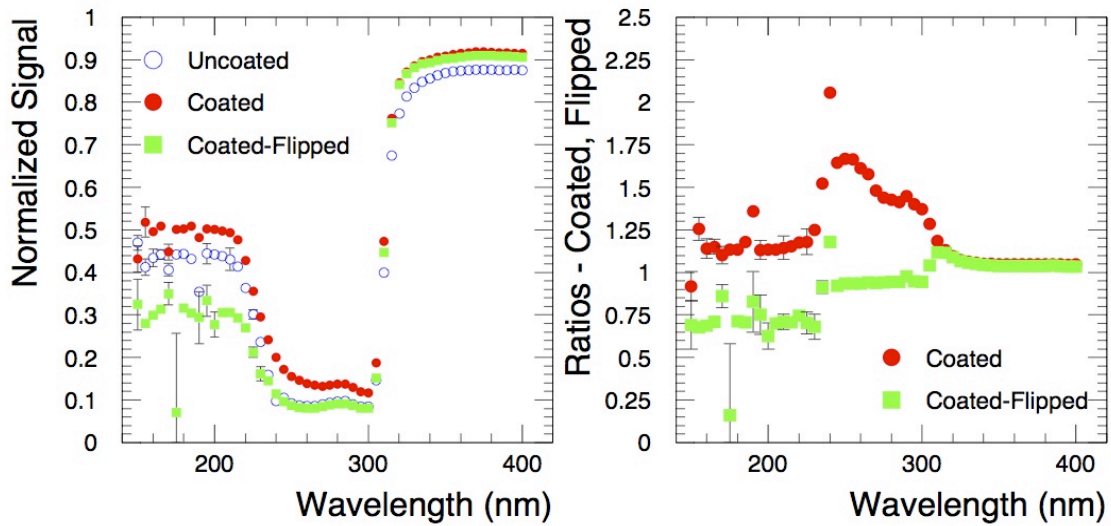

Figure 6. (Left) Response of the SiPM through plastic film only (circles), and through the plastic film with nanoparticle coating for both configurations (dots and squares), and (Right) response for Coated configuration (dots) and for Coated-Flipped (squares), each normalized by the Uncoated response.

In the left plot, the 3 signals are normalized as in the previous test to the SiPM response without the plastic film. In the right plot, ratios of the Coated/Uncoated and Coated-Flipped/Uncoated responses are shown. As expected, both ratio responses above 320 nm are the same. For all wavelengths less than 320 nm, the Coated response is greater than the Uncoated and Coated-Flipped responses. It also appears that there is a general degradation of response when more material is placed in front of the SiPM - in particular, note that the Coated-Flipped response is less than 1 for wavelengths less than ~240 nm. The feature in the Coated configuration at ~250 nm persists, possibly related to the measured response of the Si nanoparticle photodiode in [3].

## 7. Conclusion

Enhancement of the wavelength response of a Hamamatsu MPPC was observed when Si nanoparticles were deposited on a plastic film immediately preceding the photosensor. It appears that the Si nanoparticles absorbed the UV light with wavelengths as low as the $D_2$ lamp and monochromator system could deliver (>160 nm) and then emitted light at a wavelength detectable by the SiPM, turning the nanoparticle coating/photosensor combination into a UV photon detector. Further tests are now planned to determine the optimal size of the nanoparticles (for tuned wavelength response), the optimal thickness of the nanoparticle coating, and measurements of the efficiency of the response as a function of wavelength and for the SiPM operated with its normal bias voltage in Geiger mode.

## Acknowledgments

This manuscript has been created by UChicago Argonne, LLC, Operator of Argonne National Laboratory ("Argonne"). Argonne, a U.S. Department of Energy Office of Science laboratory, is operated under Contract No. DE-AC02-06CH11357. The U.S. Government retains for itself, and others acting on its behalf, a paid-up nonexclusive, irrevocable worldwide license in said article to reproduce, prepare derivative works, distribute copies to the public, and perform publicly and display publicly, by or on behalf of the Government. This work was also supported in part by US National Science Foundation grant OISE 11-03-398.

## References

[1] R. Mao et al., *Crystals for the HHCAL Detector Concept*, in proceedings of XVth International Conference on Calorimetry in High Energy Physics (CALOR 2012), *J. Phys.: Conf. Ser.* 404 (2012).

[2] H. Wenzel et al., *Simulation of a totally active dual readout calorimeter for future lepton colliders*, FERMILAB-PUB-11-531-CD-E, 2012, and S. Magill, *Use of Particle Flow Algorithms in a Dual Readout Crystal Calorimeter*, in proceedings of XVth International Conference on Calorimetry in High Energy Physics (CALOR 2012), *J. Phys.: Conf. Ser.* 404 (2012).

[3] O.M. Nayfeh et al., *Thin Film Silicon Nanoparticle UV Photodetector*, *IEEE Photonics Technology Lett.* Vol. 16, No. 8, August 2004.

[4] G. Belomoin, et al., *Observation of a magic discrete family of ultrabright Si nanoparticles, Appl. Phys. Lett.* 80, 841 (2002).

[5] D. Nielsen et al., *Current-less anodization of intrinsic silicon powder grains: Formation of fluorescent Si nanoparticles*, *J. Appl. Phys.* 101, 114302 (2007).

[6] L. Mitas et al., *Effect of surface reconstruction on the structural prototypes of ultrasmall ultrabright Si29 nanoparticles, Appl. Phys. Lett.* 78, 1918 (2001).

[7] Hamamatsu Photonics, www.hamamatsu.com/us/en/product/category/3100/4004/4113/index.html, (2014).

[8] Newport Corporation, www.newport.com/Deuterium-Lamps/378014/1033/info.aspx, (2014).